**DNA-TiO$_2$ nanoparticle nanoassembies: effect of temperature and nanoparticles concentration on aggregation**

Evgeniya Usenko[1], Alexander Glamazda[1,2*], Anastasiia Svidzerska[1], Vladimir Valeev[1], Anna Laguta[2,3],
Sergey Petrushenko[2], and Victor Karachevtsev[1]

[1]B.Verkin Institute for Low Temperature Physics and Engineering of the National Academy of Sciences of Ukraine,
47 Nauky Ave., Kharkiv 61103, Ukraine

[2]V. N. Karazin Kharkiv National University, 4 Svobody sq., Kharkiv 61022, Ukraine

[3]Aston University, Department of Mathematics, B47ET, Birmingham, UK

* Corresponding authors: glamazda@ilt.kharkov.ua, usenko@ilt.kharkov.ua

**Abstract**

TiO$_2$ nanoparticles (NPs) have unique photocatalytic properties, which are used in food industries, medicine, biosensorics, and solar energy conversion. Since the toxic properties of TiO$_2$ NPs have been insufficiently studied, additional information on the molecular mechanisms of their biological action on the structure and stability of biological macromolecules is needed, especially concerning DNA. In this work exploiting the differential UV-visible spectroscopy, the effect of the heating (from 20 till 90 $^0$C) and concentration of TiO$_2$ NPs ((1-3)×10$^{-4}$ M) on a conformation of native DNA adsorbed on TiO$_2$ NPs in a buffer solution (pH 5) is studied. Analysis of dynamic light scattering (DLS) data for the DNA:TiO$_2$ NPs suspension revealed that when the temperature increases the separated DNA:TiO$_2$ NP nanoassemblies form nanoaggregates. Correlation between the thermal dependency of the DLS data and thermal DNA denaturation measurements indicated that the appearance of the single-stranded unwound regions in double-stranded DNA in the suspension with temperature rise promotes the effective formation of the DNA:TiO$_2$ NPs nanoaggregates.

**Introduction**

Recent advances in the field of nanotechnology stimulate the large-scale production of various innovative nanomaterials. Many of them have unique physicochemical properties originating from quantum confinement effects. TiO$_2$ nanoparticles (TiO$_2$ NPs) are among the most used nanomaterials [1, 2]. They have found wide applications in the field of environmental engineering, cosmetology, pharmaceuticals, and medicine [3–7]. Despite the intensive use of TiO$_2$ NPs in various spheres of human life, many studies are informing that these NPs can pose a threat to the environment and human health [5,8-11]. However, in scientific literature, we can find contradictory information regarding the genotoxicity of these nanomaterials. In this regard, the study of the interaction between TiO$_2$ NPs and DNA is an actual topic. At present, there are already some experimental data devoted to studying the effect of TiO$_2$ NPs on DNA [12–16]. In particular, in Ref. [15], the effect of TiO$_2$ NPs on DNA was studied by the spectroscopic methods which are very sensitive to the structural modifications of DNA. It was shown that the binding of TiO$_2$ NPs to DNA is caused by both the electrostatic interaction between positively charged TiO$_2$ NPs and negatively charged DNA phosphate groups and the formation of chemical bonds between TiO$_2$ NPs and DNA [15]. The binding of DNA to TiO$_2$ NPs stimulates the disruption of the secondary structure of the biopolymer. In Ref. [16], the molecular mechanisms of DNA damage in HepG2 cells induced by TiO$_2$ NPs were studied. It was shown that TiO$_2$ NPs can affect gene expression in DNA. In Ref. [12], using data obtained by fluorescent titration, the binding constant of TiO$_2$ NPs to DNA was determined as $\sim 4.2 \times 10^6$ M$^{-1}$. The results obtained in this work showed that TiO$_2$ NPs have a high affinity for DNA and they can directly bind to the biopolymer. In addition, it was reported that TiO$_2$ NPs can strongly inhibit DNA replication and change the conformation of polynucleotides, which can lead to genotoxicity [12]. It should be noted that, in some cases, the temperature is a qualitative analog of the action in the cell of a special enzyme called helicase, which is directly involved in the unwinding of double-stranded DNA, obtaining two separated single-stranded chains and preparing them for replication. Previously, we have studied the effect of TiO$_2$ NPs on the thermal stability of DNA [17] with and without UV irradiation. However, in that study, the measurements were performed in a limited concentration range of TiO$_2$ NPs of [$c_{TiO2}$]=2.5$\times 10^{-5}$-1.5$\times 10^{-4}$ M. A temperature range in which the formation of the DNA:TiO$_2$ NP nanoassemblies (separated TiO$_2$ NPs covered by DNA) is more intense than at room temperature has been determined. The presence of the DNA:TiO$_2$ NP nanoassemblies was observed by both the TEM and DLS methods. It should also be noted that the possibility of the formation of the DNA:TiO$_2$ NP nanoassemblies was shown in Ref. [13, 18–22]. It was supposed that both single-stranded ends of double-stranded DNA and double-stranded DNA can be wrapped around TiO$_2$ NPs [13, 18]. Currently, there are works in which the adsorption of DNA on TiO$_2$NPs [13, 18–22] was observed, as well as the aggregation of the DNA:TiO$_2$ NP nanoassemblies [23, 24]. However, these studies were carried out using short fragments of DNA or RNA at fixed temperatures. There is practically no information about the study of the effect of temperature on the formation of such complexes (including complexes of TiO$_2$ NPs with natural DNA), as well as their possible aggregation with the increasing TiO$_2$ NPs concentration.

The present work is aimed to study the effect of the heating and concentration of TiO$_2$ NPs on the stability of DNA adsorbed on TiO$_2$ NPs in a buffer solution at pH 5. The melting of DNA and the injection of nanoparticles with a subsequent increase in their concentration can be accompanied by aggregation of DNA:TiO$_2$ NP nanoassemblies. Thus, in this paper we intend to identify a possible correlation between these two factors affecting the suspension of DNA:TiO$_2$ NPs.

**Materials and methods**

*Materials*

The TiO$_2$ NPs powder purchased from Sigma-Aldrich (particle diameter (d)<100 nm (BET), Product code: 677469) was used in the present work. TiO$_2$ NPs were dispersed in distilled water with pH of 6.65 which is close to the isoelectric point. The NPs suspension was ultrasonicated with $v$=22 kHz for 40 min at room temperature. The complexes of TiO$_2$ NPs with DNA were prepared as follows: 1) the salmon sperm DNA (M$_w$=(4-6)×10$^6$ Da) purchased from Serva (Germany) was added to a buffer solution with 10$^{-3}$ M sodium cacodylate (CH$_3$)$_2$AsO$_2$Na•3H$_2$O from Serva (Germany), 0.099 M NaCl at pH 5; 2) the required amount of the TiO$_2$ NPs was added to the DNA buffer suspension. The DNA phosphorous concentration [P] was (7.14±1)×10$^{-5}$ M that had been determined by the molar extinction coefficient at $v_m$=38500 см$^{-1}$ [25]. In the present work, DLS study of the DNA:TiO$_2$ NP nanoassemblies was performed in the cacodylate buffer suspension (0.1 M Na$^+$, pH 5) at [c$_{TiO2}$]=1.5×10$^{-4}$ M. The concentration of polynucleotide phosphates [P] equalled 8×10$^{-5}$ M. The error in determining the concentration of TiO$_2$ NPs and polynucleotide phosphorus did not exceed 0.5 %.

*Transmission electron microscopy (TEM)*

A visualization of TiO$_2$ NPs and DNA:TiO$_2$ NPs nanoaggregates was realized using transmission electron microscopy (SELMI EM-125, Sumy, Ukraine). For in situ electron diffraction studies, samples were sequentially condensed into fresh cleavages of KCl single crystals with an amorphous carbon sublayer deposited on them. A 3-µL sample drop was deposited and adsorbed for 1 min, and then, the excess of the suspension was removed with a piece of filter paper.

*Dynamic light scattering*

The particle size distribution of suspended DNA:TiO$_2$ NP nanoassemblies was measured using the dynamic light scattering (DLS) method. The hydrodynamic size of nanoparticles consists of the real diameter (according for example TEM measurements) plus the double thickness of the double electric layer. Thus the hydrodynamic size exceeds the real (core) size of the nanoparticles. The particle size distributions were determined via DLS using Zetasizer Nano ZS (Red badge) ZEN 3600 Malvern Instruments apparatus in the temperature range of 25-90 $^0$C. To measure the diffusion speed, the speckle pattern was produced by illuminating the particles with a 4 mW 632.8 nm He-Ne gas laser and 175$^0$ detection optics allowing analyze the scattering information at close to 180$^0$. The laser power was automatically attenuated so that the count rate from the sample, especially high scattering samples, was within acceptable limits. The system was heated automatically in a cuvette holder with a temperature controller with a step 2 $^0$C. The equilibration time was 30 s. This time interval didn't include the short time for temperature regulation by the apparatus. There are 3 particle size measurements for each temperature point (10 in the case of 25 $^0$C). Every measurement consisted of at least 10 runs (automatic choice). To assign the viscosity value was used "solvent builder", which calculated the viscosity of the dispersant depending on NaCl concentration. Glass cuvette was used for size measurements. The result of every run is a correlation curve [26]. Every measurement is based on an average value of the correlation curves of all runs. Such averaging correlation curve was treated by two mechanisms [27]. The first one is a cumulant analysis which is the fit of a polynomial to the log of the correlation function and the second order cumulant is the Z-average diffusion coefficient. We obtained the Z-average diameter (this is hydrodynamic size) by this method but it is good fitting only when one analysis for spherical, reasonably narrow monomodal samples, i.e. with polydispersity below a value of 0.1. The polydispersity index characterizes the uniformity of colloidal particles (one or more sizes of particles are presented in the sample) and, subsequently, we can use it as a system quality index. As a quality index, we also used standard deviation, calculated by Zetasizer software for every temperature position. The second mechanism of the treatment of correlation curves is that the correlation function is a superposition of exponential decay for each particle size in the sample. Then coefficients of exponential decay can be transformed to particle size distribution (CONTIN algorithm). It should be noted that in general DLS technique is insensitive to the particular size distribution (to separate peaks with some size) [27]. Moreover, the particle size distributions calculated from coefficients of exponential decay are the distributions by the intensity of scattering light. However, Zetasizer software allows calculating using Mie theory two other distributions – by the volume and by the number of scattering particles – since the intensity of scattering light is proportional to the particle diameter to the 6th power, the volume is proportional to the particle diameter to 3rd power and the number is proportional to the particle diameter [28]. Consequently, we obtained three distributions by intensity of scattering light, by the volume of scattering particles and by the number of scattering particles (from measuring the fluctuation in scattering intensity), as well as hydrodynamic size (Z-ave or $d_H$), which is calculated from diffusion coefficient (cumulant analysis). To choose the appropriate one for a given system it should be noted that the European Commission defined that nanomaterials should be characterized as nanoscale objects in terms of the number size distribution of their constituent particles [29-31]. So, it seems reasonable that we need to use the DLS size distribution by the number. Moreover, by TEM methodology we obtained the distribution of nanoparticles size by the number of these particles [32,33]. So, we can compare only distributions by the number from DLS and TEM results. It should be noted that DLS measurements are carried out in the material's native environment (colloidal solution) but in the case of TEM measurements nanoparticles were precipitated from the aqueous solutions. Such a preparation method can alter or destroy the sample [34]. Finally, exactly the distribution by the number was used for the investigation of TiO$_2$ NPs aggregation processes as reported earlier [35].

*Differential UV and visible spectroscopy*

It is known that the DNA light absorption within the ultraviolet range (UV) of 30000-50000 cm$^{-1}$ is caused by $\pi{\rightarrow}\pi^*$ and n$\rightarrow\pi^*$ electronic transitions in the nucleobases [36]. The change in the conformation and structural stability of nucleic acids occurs under the influence of various factors (for example, temperature, ionic conditions, UV exposure, etc.) and causes changes in the UV absorption spectrum [37]. The extinction coefficient of a double helix depends on the mutual orientation of the intrinsic dipole moments of stacking nitrogenous bases. If the moments of two stacked bases are collinear, the light absorption will decrease (so-called hypochromism). But the disordering of the stacked biopolymer structure induces the delocalization of the moments and the increase in the light absorption (so-called hyperchromism). Differential UV (DUV) spectroscopy is used to reveal the weak changes in the electronic structure of DNA induced by environmental influence. DUV spectra of the DNA:TiO$_2$ NPs suspensions were recorded with Specord UV-vis spectrophotometer (Carl Zeiss, Jena, Germany) utilizing the four-cuvette scheme [38]. Cuvette #1 containing the buffer suspension of DNA and cuvette #3 with the buffer solution were placed in the sample channel of the spectrophotometer. Cuvette #2 with identical to cuvette #1 DNA in the buffer suspension and cuvette #4 with identical to cuvette #3 the buffer solution were placed in the reference channel. DUV were measured after the addition of the same TiO$_2$ NPs amount into cuvettes #1 and #4. To compensate for the DNA dilution in cuvette #1, the appropriate amount of the buffer was added to cuvette #2. The spectra were normalized to the concentration of polynucleotide phosphates as follows: $\Delta\varepsilon(v)=\Delta A(v)/[P]$, where $\Delta\varepsilon(v)$ is the change in light absorption at T=T$_0$ and $\Delta A(v)$ is the change in the optical density of the solution at T=T$_0$. The light power density of the radiation source was quite low which did not lead to a noticeable degradation of the samples. The taken optical spectra were measured when suspensions were in equilibrium, as evidenced by their unchanged shape and intensity for 20–30 min. The deuterium electrical lamp was used as the UV radiation source and the tungsten halogen one was employed as a source of UV-visible optical ranges.

*Thermal denaturation*

The thermal measurements of the DNA spectral peculiarities are important for the study of the possible structural biopolymer conformations. The performed DNA spectral analysis allows determining some thermodynamic binding parameters of ligand to DNA, such as the melting temperature, the temperature of the beginning and end of the helix-coil transition, and the transition interval. In the present work, DNA thermal denaturation was used to study the structural stability of the biopolymer in the presence of $TiO_2$ NPs. Upon increasing the temperature, the ordered structure of DNA is disrupted. As a result, the biopolymer forms loops which appear because of breaking some H-bonds between the nitrogen paired bases. When the temperature reaches a value close to the melting temperature ($T_m$), the number of helical regions becomes equal to the number of unwound regions. In other words, $T_m$ is the temperature at which 50% of a DNA sequence is in the helix conformation, and the other 50% is present as single strands. A further increase in the temperature leads to a strong shift of equilibrium towards the increase of the fraction of the single strands. This process is accompanied by an increase in the UV absorption intensity. The evolution of this increase stops when the double helix of the biopolymer is completely unwound. The melting curve describes the dependence of the UV absorption intensity on temperature. The melting curves of DNA with different $TiO_2$ concentrations were recorded using UV-spectrophotometer at a fixed wavenumber of $v_m$=38500 см$^{-1}$ ($\approx$260 nm) that corresponds to the maximum absorption of DNA. We have used the laboratory software allowing to perform the registration of the melting curves as the temperature dependence of the hyperchromicity coefficient: $h(T)=[\Delta A(T)/A_{T_0}]_{v_m}$, where $\Delta A(T)$ is a change in the optical density of the DNA suspension upon heating and $A_{T_0}$ is the optical density at $T=T_0$. Thus, $h(T)$ is the quantitative characteristic of hyperchromism. The registration of the absorption intensity was carried out using the double-cuvette scheme: identical suspension of DNA or their complexes with $TiO_2$ were placed in both channels of the spectrophotometer. The reference cuvette was thermostated at $T=T_0\pm0.5°C$, while the sample cuvette was slowly heated at a rate of 0.25°C/min from 20 to 94°C. The temperature in both cuvettes was determined with an error not exceeding 0.5 $°C$. In our experiments $T_0$, was equal to 20°C.

**Results and Discussion**

In our measurements TEM was used to examine the particle size and morphology of the $TiO_2$ NPs distribution. TEM image of $TiO_2$NPs sample is shown in Fig. 1a. It is seen that the $TiO_2$NPs consists of mostly spherical nanoparticles. The mean diameter of the $TiO_2$ nanoparticles estimated from taken TEM images was approximately 74 nm which is close to the declared nominal nanoparticle dimension by the manufacturer. Figure 1b shows TEM image of nanoaggregates consisting of the DNA:$TiO_2$ NP separated nanoassemblies which were deposited on the substrate from the buffer solutions. Their image with higher magnification is shown in Fig. 1c. Now we cannot answer the question, of whether the DNA:$TiO_2$ NPs nanoaggregates have been formed in the aqueous suspension and then deposited on the surface or they are the results of the adhesion of the individual nanoassemblies after drying water. The study of this problem plays an important role, for example, in the development of drug-related products based on $TiO_2$ NPs, where separation and size control of $TiO_2$ NPs is necessary. Let us now turn to the analysis of the spectroscopic studies of the DNA:TiO$_2$ NPs suspension. It's well known that spectroscopy is very sensitive to the aggregation process and with its help it's possible to reveal the minor changes in the dynamics of the aggregation.

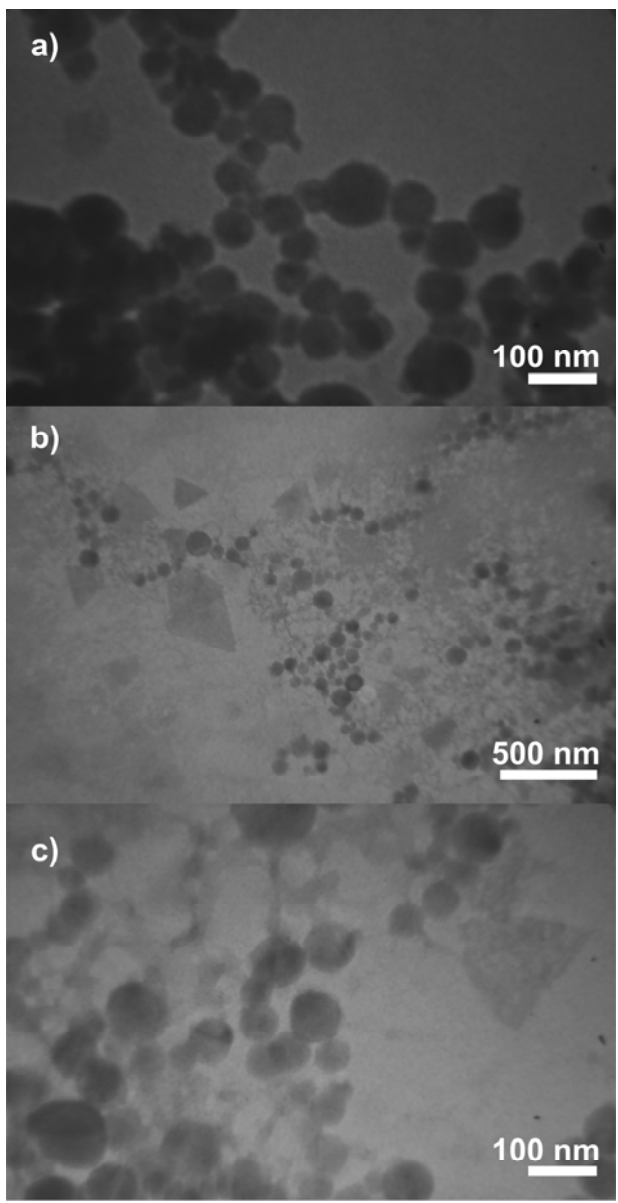

**Fig. 1** TEM images of **a)** TiO$_2$ NPs material precipitated from aqueous solution; **b)** and **c)** DNA:TiO$_2$ NPs nanoaggregates at different magnifications.

Figure 2 shows DUV absorption spectra of the DNA:TiO$_2$ NPs suspension demonstrating a stepwise evolution depending on the TiO$_2$ NPs concentration. The spectra can be visually separated into two spectral ranges at about 410 nm: 1) the low-energy range ($\lambda \geq 410$ nm), which is characterized by hypochromism; 2) the high-energy part of the spectrum ($\lambda \leq 410$ nm) demonstrating hyperchromism. It's well known that the UV absorption spectrum of DNA is limited by roughly 300 nm. The estimated band gap of the TiO$_2$ NPs used in this work is about 422 nm (2.94 eV) [17]. Thus, the observed light absorption in the visible range is mainly due to the light scattering phenomenon. The significant response observed in the UV absorption spectra is caused by both electronic absorption and light scattering in the studied system. Based on the above-mentioned information, we believe, that the presented DUV absorption spectra in Fig. 2 can be assigned to the optical characteristics of both the individual constituent elements and bound DNA:TiO₂ NP nanoassemblies. The entangled evolution of the DUV spectra reflects critical conformational changes in the DNA structure upon an increase of [$c_{TiO2}$], indicating the strong effect of TiO₂NPs on the biopolymer structure [12]. According to [12], the observed hyperchromism in the high-energy range is primarily due to a violation of the secondary structure of DNA. The UV spectral contribution caused by the absorption and scattering of light by TiO₂ NPs is more compensated by using the reference cuvette with the TiO₂ NPs suspension in the four-cuvette scheme. The part of the spectrum above 410 nm, as we suppose, can be primarily related to the light scattering by TiO₂ NPs. The increase in the TiO₂ NPs concentration can stimulate the growth of the nanoaggregates in the reference cuvette containing the TiO₂ NPs suspension [39,40]. We believe that the observed hypochromism is caused by the faster growth of the TiO₂ NPs nanoaggregates (located in the cuvette of the reference channel) than both the light scattering DNA:TiO₂ NP nanoassemblies and nanoaggregates (located in the sample channel). The inset in Fig. 2 shows the dependence of the normalized integrated intensities of the high-energy ($I_H$) and low-energy ($I_L$) parts of the DUV spectra separated by the point at about 410 nm and bound by the $\Delta\varepsilon=0$ line. Both dependences $I_H$ and $I_L$ demonstrate the quasi-linear behavior in dependence on [$c_{TiO2}$]. However, there is a clear inflection point on the $I_H/I_L$ curve near [$c_{TiO2}$]≈2.25×10⁻⁴ M. We suppose that sudden change in the concentration dependence may be explained by the intensification of the formation of the TiO₂ NPs nanoaggregates in the reference cuvette with the TiO₂ NPs suspension (Fig. 2, inset). In another cuvette, we suppose, the biopolymer covering TiO₂ NPs slows the formation of DNA:TiO₂ NPs nanoaggregates at room temperature.

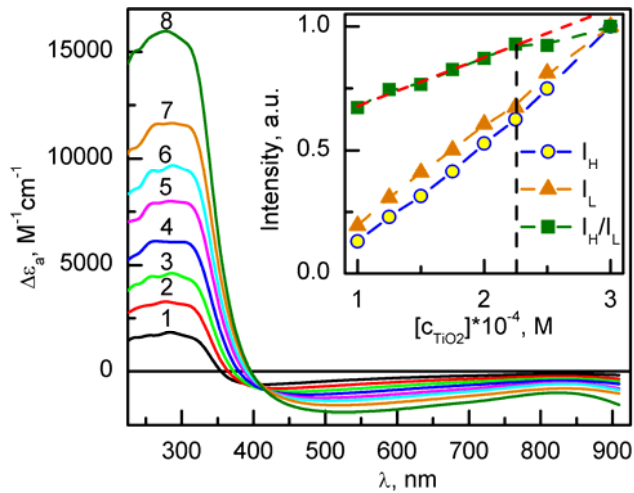

**Fig. 2** DUV spectra of the DNA:TiO₂ NPs suspensions at T=25⁰C: «1» - 10⁻⁴ M; «2» - 1.25×10⁻⁴ M; «3» - 1.5×10⁻⁴ M; «4» - 1.75×10⁻⁴ M; «5» - 2×10⁻⁴ M; «6» - 2.25×10⁻⁴ M; «7» - 2.5×10⁻⁴ M; «8» - 3×10⁻⁴ M. The inset shows the dependence of the normalized integrated DUV spectral intensities of the high-energy ($I_H$), low-energy ($I_L$) parts and the ($I_H/I_L$) ratio as the function of [$c_{TiO2}$]. The $I_H/I_L$ curve is fitted with linear function (red dashed line). The anomaly near [$c_{TiO2}$]≈2.25×10⁻⁴ M on the $I_H/I_L$ curve is shown with a vertical dashed black line.

The typical temperature dependence of the hyperchromic coefficient (h) of DNA has the S-like shape: heating induces the helix-coil structural transition, as reflected in the demonstration of the hyperchromism (h>0) [17]. The analysis of the DNA melting curve without TiO₂ NPs allowed to determine the thermodynamic parameters of the biopolymer: $T_m$=78.5 ⁰C and $h_{m0}$=0.41 (curve 1 in Fig. 3) (where $h_{m0}$ - the value of the hyperchromic coefficient in the absence of TiO₂ NPs). However, the addition of TiO₂ NPs to the DNA buffer solution leads to a dramatic change in the shape of the biopolymer melting curve and the appearance of the trough in the temperature range of about 30-75 °C [17]. The depth of the trough increases with the $TiO_2$NPs concentration. The appearance of the trough is an essential feature that the bound DNA with $TiO_2$ NPs absorbs the light less efficiently than the free native DNA in the buffer solution. As [$c_{TiO2}$] increases, several segments can be clearly distinguished on the melting curve. Figure 3 shows the analysis of the DNA melting curve in the presence of $TiO_2$ with [$c_{TiO2}$]=1.5×10$^{-4}$ M [17]. The characteristic points: $T_{s1}$ and $T_{s2}$ as well as $T_{f1}$ and $T_{f2}$were calculated using an extrapolation of the linear parts of the melting curves. They are connected to the start ($T_{s1}$, $T_{s2}$) and finish ($T_{f1}$, $T_{f2}$) of the definite ongoing processes discussed below. In the segment of a-b, with the increase in temperature, hypochromismis observed, the value of which increases with [$c_{TiO2}$] [17]. It is known that heating leads to the formation of unwound regions at the ends as well as on the stretches located far from the biopolymer ends. In this regard, it is possible to assume two main mechanisms of binding of NPs with DNA could coexist in the segment of a-b. The first type of the binding is electrostatic interaction of positively charged $TiO_2$ NPs with negatively charged phosphate groups of DNA [15]. The second type of complex could be formed by binding $TiO_2$ NPs with unwound single-stranded DNA ends which can wrap around $TiO_2$ NPs [17]. In this case, the interaction with the nitrogenous bases of DNA can play a significant role in the formation of this type of complex.

The nature of the observed hypochromism is unclear and can be a subject of further studies. It's well known that the main cause of hypochromism is the formation of an ordered DNA structure. We believe that the revealed hypochromism in our experiment can be interpreted in terms of the appearance more ordered structure of the biopolymer bound to $TiO_2$ NPs. It should be noted that we observed a similar effect earlier in the formation of a three-stranded polynucleotide from a double-stranded polynucleotide in the presence of divalent metal ions [41], as well as in the case of the formation of a metalized form of DNA [42].

In the segment of b-c, h remains constant. The denaturation of the biopolymer structure occurs in the c-d segment and ends with the plateau (Fig. 3).

With a further increase in [$c_{TiO2}$], the biopolymer melting curves undergo strong changes. In particular, the absorption hypochromism is replaced by hyperchromism at [$c_{TiO2}$]=1.75×10$^{-4}$ M which can be explained by the formation of the DNA:$TiO_2$ NPs nanoaggregates by binding the neighboring DNA:$TiO_2$ NP nanoassemblies and the appearance of the visible sediment at above [$c_{TiO2}$]=2×10$^{-4}$ M (curves 3 and 4 in Fig. 3, inset). It should be noted that the visible sediment was not observed in the suspension at room temperature and it emerged upon heating. It can be explained by an increase in the appearance of single-stranded fragments in native DNA upon heating. They can bind closely arranged nanoassemblies.

The generation of the DNA:$TiO_2$ NPs nanoaggregates in the suspension has prompted us to perform the temperature-dependent DLS measurements of the DNA:$TiO_2$ NPs suspension with [$c_{TiO2}$]=1.5×10$^{-4}$ M. The performed analysis of the melting curves allows to suppose that both nanoassemblies and nanoaggregates can coexist at this concentration of $TiO_2$ NPs. The abovementioned alternative method for estimating the size of the formed nanoaggregates in the suspension has a number of restrictions caused by the experimental technique, sample preparation protocols and sample concentration, etc. DLS method is very sensitive to the aggregation processes occurring in the suspension and it's able to probe the kinetic evolution of the formation of nanoassemblies/nanoaggregates, as shown in the studies of the thermally responsive polymers based on poly (N-isopropylacrylamide) with incorporated gold nanoparticles [43].

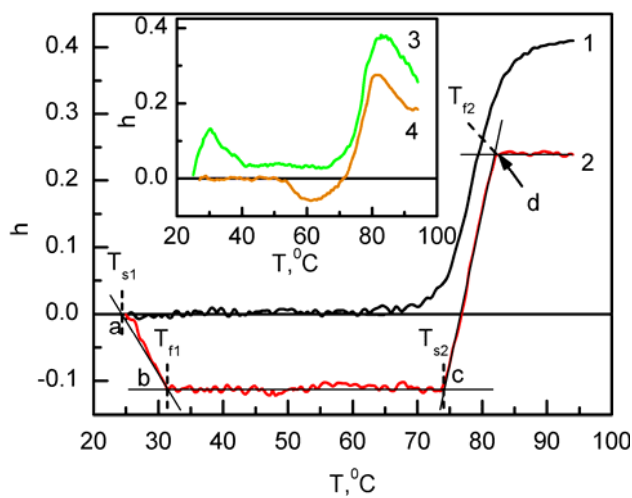

**Fig. 3** Temperature dependence of hyperchromic coefficient of DNA without (curve 1) and with (curves 2-4) $TiO_2$ NPs: «1» - [$c_{TiO2}$]=0; «2» - [$c_{TiO2}$]=1.5×10$^{-4}$ M. The inset shows the temperature dependence of hyperchromic coefficient of DNA taken at [$c_{TiO2}$]=1.75×10$^{-4}$ M (curve 3) and [$c_{TiO2}$]=2×10$^{-4}$ M (curve 4).

DLS measurement analysis of the DNA:$TiO_2$ NPs suspension performed in the temperature range of 25-90 $^0$C revealed the size dependent evolution of the scattering nanoassemblies. Figure 4 shows the size distribution map of the DNA:$TiO_2$ NP nanoassemblies (obtained from the distribution curves by the number) taken from 25 to 90$^0$C. There is one nanoassembly size distribution cone with a maximum at about 90 nm observed at room temperature. Upon heating up to about 30$^0$C, the cone undergoes splitting into two ones with peaks at 60 and 220 nm. So, we have a situation when there are a small number of large DNA:$TiO_2$ NPs nanoaggregates and a large number of individual DNA:$TiO_2$ NP nanoassemblies or small nanoaggregates. The intensity of the first cone at 60 nm demonstrates consistent suppression and, conversely, the second cone shows an increase in the intensity upon heating. It seems reasonable, that more nanoaggregates with a mean diameter of about 220 nm appear upon heating. The threshold redistribution of the intensities of two cones is observed at about 50$^0$C. The intensity of the cone at about 60 nm is almost suppressed at 70$^0$C and vanished at 85$^0$C. In turn, the intensity of the cone at 220 nm is raised upon heating. It reaches a maximum at about 70$^0$C and decreases up to 80$^0$C. And then after that, the intensity of the cone at 220 nm begins to increase again. Now let's look closely at the size distribution of the DNA:$TiO_2$ NP nanoassemblies taken at 25 and 90 $^0$C. Figure 5 shows that the diameter of the DNA:$TiO_2$ NP nanoassemblies increases by 2.5 times upon heating. It may be explained by intensifying the formation of single-stranded regions in DNA upon heating. These single-stranded regions are much more flexible than double native DNA. It can promote more efficient binding of the single-stranded regions to nanoparticles than the double-stranded ones. Upon heating and unwinding the native DNA the length of single-strand sections increases and DNA strands can bind two or more nanoassemblies into the nanoaggregates.

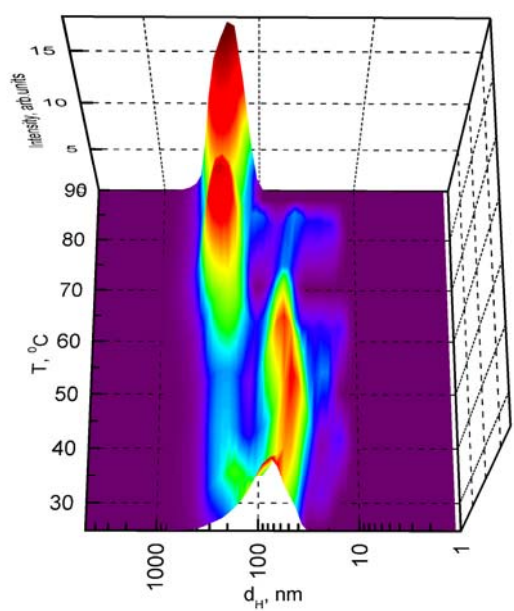

**Fig. 4** The size distributions map by the number for DNA:TiO₂ NP nanoassemblies colloidal suspension at pH 5 (cacodylate buffer) at the different temperatures, T=25-90 $^0$C. The surface was built after averaging of the distributions in the range of every 10 degrees.

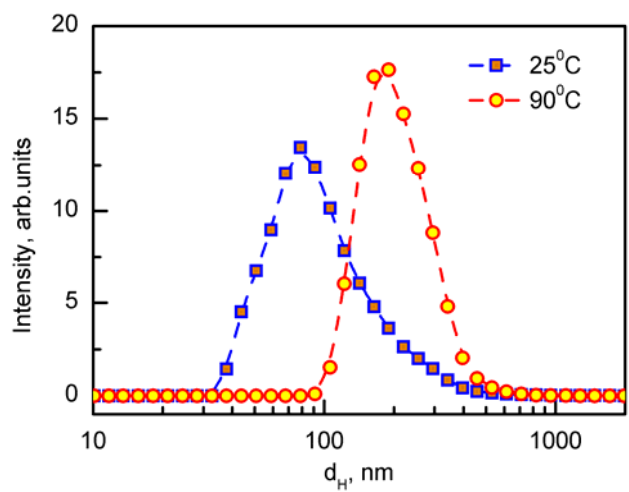

**Fig. 5** The size distributions by the number for the DNA:TiO₂ NP nanoassemblies colloidal suspension at pH 5, T=25 and 90 $^0$C. Indexes of polydispersity are PdI = 0.34±0.05 (25 $^0$C), PdI = 0.24±0.01 (90 $^0$C). The mean number size (from Malvern Zetasizer Software) is 109±44 (25 $^0$C) and 213±8 (90 $^0$C). The experimental and calculated correlation functions are plotted in Supplementary Information file.

Summing up the spectral studies we can conclude that the DNA:TiO₂ NP nanoassemblies are generated at room temperature, and the formation of their nanoaggregates is intensified upon heating as a result of unwinding of native DNA. We suppose, that the DNA:TiO₂ NPs nanoaggragates presented in the TEM image (Fig.1b, c) can be formed after the deposition on the surface and drying of water.

It should be noted that the study of colloidal solutions of biopolymers with NPs, as well as the determination of the sizes of the formed nanoassemblies/nanoaggregates, is a rather complex experimental task that requires a comprehensive approach.

**Conclusions**

For the first time, the influence of temperature and TiO₂ NPs concentration on aggregation of DNA:TiO₂ NPs nanoassemblies was studied by UV-spectroscopy and thermal denaturation.

The study of the thermal stability of DNA suspensions in the presence of high concentrations ($\geq 1.5 \times 10^{-4}$ M) of TiO₂ NPs showed that with the injection of these NPs the formation of large nanoaggregates consisting of several DNA:TiO₂ NP nanoassemblies occurs. It had been shown that heating intensifies the formation of DNA:TiO₂ NPs nanoaggregates that correlates with performed DLS studies. This showed that upon heating from room temperature up to about 30 $^0$C, the observed cone at 90 nm undergoes splitting into two ones with peaks at 60 and 220 nm. Upon heating the number of the DNA:TiO₂ NPs nanoaggregates with a mean diameter of about 220 nm growths due to the appearance of single-stranded regions in native DNA which can more effectively bind TiO₂ NPs to nanoaggregates than double-stranded regions. The results obtained are fundamental, and can also be used to create self-cleaning antibacterial surfaces, as well as in medicine. We suppose that the presented results in the given work will stimulate and facilitate further studies on complex biological systems with inorganic nanomaterials.

**Acknowledgments**

Authors acknowledge financial support from National Academy of Sciences of Ukraine (Grant No. 0120U100157). The authors gratefully thank Dr. Alexander Plokhotnichenko and Dr. Alexey Peschanskii (B.Verkin Institute for Low Temperature Physics and Engineering of the National Academy of Sciences of Ukraine) for the interest in the presented work and the valuable discussions.

**Ethics approval and consent to publish, participate**

We confirm that this manuscript had not been published elsewhere before and would not be considered to be published on other journals.

**Conflicts of Interest**

The authors do not have any commercial or associative interest that represents any conflict of interest in connection with the work submitted.

**Author contribution statement**

All authors discussed the results and commented on the manuscript. V. Valeev and E. Usenko carried out the spectroscopic measurements, thermal denaturation and analyzed data. A. Svidzerska and A. Laguta performed and analyzed the DLS data. S. Petrushenko performed TEM characterization. A. Glamazda and V. Karachevtsev planned and coordinated the project. E. Usenko, A. Glamazda, and V. Karachevtsev wrote the paper.